\documentclass[conference]{IEEEtran}
\IEEEoverridecommandlockouts
\usepackage{cite}
\usepackage{amsmath,amssymb,amsfonts}
\usepackage{algorithmic}
\usepackage{graphicx}
\usepackage{textcomp}
\usepackage{xcolor}
\def\BibTeX{{\rm B\kern-.05em{\sc i\kern-.025em b}\kern-.08em
    T\kern-.1667em\lower.7ex\hbox{E}\kern-.125emX}}

    \UseRawInputEncoding

\begin{document}

\title{ Power Evaluation of IOT Application Layer Protocols \\
}

\author{\IEEEauthorblockN{1\textsuperscript{st} Amirhossein Shahrokhi}
\IEEEauthorblockA{\textit{Computer and Information Technology Department } \\
\textit{Razi University}\\
Kermanshah, Iran \\
amir.shahrokh640@gmail.com}
\and
\IEEEauthorblockN{2\textsuperscript{nd} Mahmood Ahmadi}
\IEEEauthorblockA{\textit{Computer and Information Technology Department} \\
\textit{Razi University}\\
Kermanshah, Iran \\
m.ahmadi@razi.ac.ir}

}

\maketitle

\begin{abstract}
The Internet of Things has affected all aspects of daily life, and the number of IoT devices is increasing day by day. According to forecasts, the number of Internet of Things devices will reach one trillion devices by 2035. The increase in the number of devices connected to the Internet will cause various concerns. One of the most important concerns is the energy and power consumption of these devices. Although Internet of Things modules are low in energy consumption, their widespread and large-scale use has made the issue of power consumption become the most important challenge in this field. For this reason, it is necessary to use communication protocols that, in addition to establishing efficient communication, impose minimal power consumption on the network. In this paper, application layer protocols such as MQTT, MQTT-SN, CoAP, and HTTP are simulated using the tools available in the Contiki operating system, including COOJA and Powertrace, and they { are evaluated} and compared with each other in terms of power consumption. According to the simulations performed by the mentioned tools, the MQTT-SN protocol was the least consuming protocol in terms of power consumption. After that, the CoAP protocol is placed, and with a slight difference, the MQTT protocol, which consumes more than MQTT-SN. Finally, the HTTP protocol consumes the most power, which makes it unsuitable for communication in the Internet of Things

\end{abstract}


\section{Introduction}
\label{intro}
The increase in the number of devices and the number of connections in the field of Internet of Things has caused the discussion of the consumption power of these devices to become a very important and vital issue. The communication of these devices in this field is formed under certain rules and protocols, like web communication, so that they provide a platform for establishing communication between devices that will be working in different layers of the network \cite{ref12}. These devices work by using sensors that transmit data to a computer or software and allow them to perform important tasks. Due to their application and high level of automation, the number of devices that connect to the Internet has increased \cite{ref13}. On average, 127 objects are synchronized per second. Connecting different objects together has many advantages. {Below are some of the benefits: the possibility of tracking and monitoring devices, more data means better decisions, lower workload, increase efficiency with savings, and better quality of life.}

Creating more energy-efficient technologies with anticipated deployments everywhere (the so-called Internet of Things or Industrial Internet) that enables optimal industrial exploitation and helps improve social welfare. Modeling and measuring the energy consumption of an application in pre-deployment or pre-production stages is of great importance due to the critical requirements of IoT applications in terms of cost reduction, lifetime and available energy. Optimizing energy consumption is a key factor for scalable IoT products and an important IoT design consideration, especially for developers of battery-powered devices such as smart city applications, and routing and tracking solutions. To make business sense, IoT devices must work continuously and reliably in this field for about 10 years \cite{ref14}. 

One of the things that causes energy consumption and power increase in the Internet of Things world is the use of inappropriate communication protocols in different layers. In this research, the goal is to evaluate the power consumption of application layer protocols using COOJA simulation tools on a virtual platform. The main contribution of this paper is as follows:
\begin{itemize}
\item Running the Cooja simulator in the Contiki operating system in a virtual environment and then running Internet of Things application layer protocols such as MQTT-SN, MQTT, COAP and HTTP.
\item Using the Zolteri-Z1 module, which is known as Z1-mote in the Cooja emulator, and then using the Powertrace tool to evaluate the power consumption of the mentioned protocols.
\end{itemize}
The rest of this paper is organized as follows. Section \ref{related} presents the related work on the evaluation of IoT  application layer protocols. Section \ref{protocols} describes the summarized concept about the application layer protocols in IoT. Section \ref{proposed} represents the proposed method for the evaluation of these protocols. Section \ref{results} describes the evaluation results and section \ref{conclusion} concludes the paper.

\section{Related Work}
\label{related}
In this section, the related researches in IoT application layer protocols {are} reviewed.

In \cite{ref3}, an analysis of the resources required by CoAP and MQTT protocols is done based on experimental results using Intel X86 processors with Ubuntu operating system, libcoap and Mosquitto libraries. They are compared based on the specific features of different protocols as well as the use of resources such as bandwidth and energy. The analysis of experimental results shows that CoAP is more efficient in terms of energy consumption as well as bandwidth.

In \cite{ref4}, a quantitative and qualitative comparison between MQTT and CoAP when used in a smartphone application is presented. For this study, they use a CoAP server and client based on Java implementation called Californium. The MQTT broker is based on Mosquitto. MQTT publisher and CoAP server run on a smartphone with Android 4.2.2, and MQTT {broker} and CoAP client run on a laptop with Windows 7 operating system. They measure bandwidth usage and round-trip time and conclude that CoAP can be a valid alternative to MQTT for certain application scenarios.
In \cite{ref5}, the costs of using CoAP and HTTP in IoT applications are analyzed based on a theoretical cost model. The analysis shows that there are simpler hardware requirements in CoAP, as well as lower communication costs and reduced energy consumption in this protocol. As a result, it leads to cost reduction in many application scenarios. CoAP requires only 40\% of the power typically used by HTTP.
In \cite{ref6}, CoAP and MQTT-SN have been evaluated for robot communication. For the experiments, they use a Raspberry Pi B (ARMv6 processor) connected to the network via LAN Ethernet cable, using CON and QoS1. They only measure the transfer {time} and find that MQTT-SN shows 30\% better transfer times. 
In \cite{ref8}, the efficiency, usage and requirements of MQTT and CoAP are discussed and analyzed using Raspberry-Pi with Raspbian operating system and temperature sensor in a simple experiment that includes a publisher, server and broker. They conclude that CoAP handles more data than MQTT as the size of the sent message increases.
In \cite{ref11}, they look at the effect of application layer {protocols} (CoAP and MQTT) on performance on two different wireless networks: Bluetooth Low Energy and Wi-Fi. The evaluation was done using Ericsson's internal event-based radio network system simulator. They found that CoAP performs better both in terms of latency and energy consumption in both wireless networks. In this paper, the power consumption of application layer protocols (MQTT, MQTT-SN, COAP and HTTP) using COOJA simulation tools on Contiki operating system is evaluated.
{
In \cite{ref25}, an evaluation HTTP, MQTT, DDS, XMPP, AMQP, and CoAP protocols have been presented, and subsequently, the power consumption prediction based on linear regression model of MQTT and COAP is analyzed in order to enhance data communications in IoT applications.

In \cite{ref26}, power consumption of application layer protocols CoAP, MQTT and XMPP for the Internet of Things is analyzed. Based on the results, MQTT and CoAP provide major energy savings, unlike XMPP which consumes more power.}

\section{IoT Application Layer Protocols}
\label{protocols}
In this section a brief introduction of the MQTT, MQTT-SN, COAP and HTTP protocols are described.
\subsection {MQTT protocol}
MQTT (Message Queuing Telemetry Transport) is a publish/subscribe messaging protocol designed for M2M-style communication over limited networks. An MQTT client publishes messages to an MQTT broker that are subscribed to by other clients or may be saved for future subscription. Each message is published to an address that is known as the topic. Clients can subscribe to multiple topics and receive any message published for each topic. 

\begin{figure}[htbp]
\centerline{\includegraphics[scale=0.5]{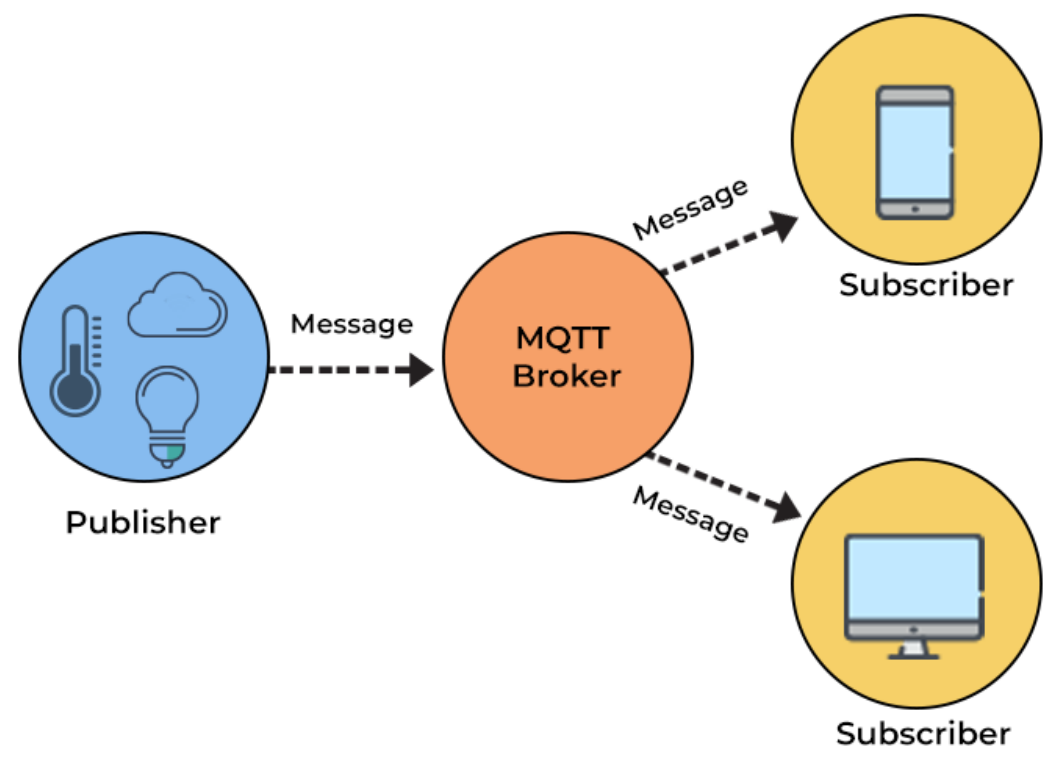}}
\vspace{-0.25cm}
\caption{An overview of MQTT protocol.}
\label{fig1}
\end{figure}

It is a full duplex protocol and typically requires a fixed 2-Byte header with a small message payload up to a maximum size of 256 MB. It uses TCP as the transport protocol and TLS/SSL for security. Therefore, the relationship between the client and the broker is communication-oriented. An overview of MQTT process is depicted in Figure \ref{fig1}.
Depending on the capabilities of the processor, MQTT can connect thousands or even millions of devices. In addition to scalability, it is simple to use.
\subsection {MQTT-SN protocol}
MQTT-SN is a protocol specifically designed for very low power M2M devices. MQTT is a protocol designed for M2M style protocol, but requires TCP-IP to work. Although MQTT is claimed to be a lightweight protocol, it is not suitable for sensors and devices that cannot maintain their own TCP-IP stack. MQTT-SN is specific to sensor networks and does not depend on TCP-IP to work. It can also work on any transport layer such as ZigBee.
The main difference between MQTT and MQTT-SN is as follows.
\begin{itemize}
\item  MQTT-SN uses a 16 bits { topic-ID} without any registration instead a Topic in MQTT. The short topic-ID assists the bandwidth constraint sensor networks.
\item MQTT-SN can change the content at any time or even cancel it. MQTT only allows content to be set during CONNECT and no changes are allowed. In the MQTT-SN network, there is a gateway device that is responsible for converting MQTT-SN to MQTT and communicating with the MQTT server in the cloud.
\item It also supports the sleep function of the device. If the device enters sleep mode and cannot receive UDP data, it sets the downstream PUBLISH gateway. This message is buffered and sent to wake up the device. 

\end{itemize}
\subsection {COAP protocol}

COAP is a lightweight M2M protocol from the IETF CoRE (RESTful Constrained Environments) working group. CoAP is a REST-like request/response protocol. CoAP is mainly used to interoperate with HTTP and RESTful Web through proxies.
Unlike MQTT, CoAP uses Universal Resource Identifiers (URIs) instead of topics. The publisher publishes data in the URI, and the subscribers subscribe to the specific resources indicated by the URI. When a publisher publishes new data to a URI, all subscribers are notified of the new value represented by the URI. CoAP uses UDP as transport protocol and DTLS for security. Therefore, clients and servers communicate through connectionless datagrams with less reliability. However, it uses "Confirmable" or "Non-confirmable" messages to provide two different levels of service quality. Confirmable messages must be acknowledged by the receiver with an ACK packet, and non-confirmable messages are not.

\section{Proposed Method}
\label{proposed}

In this paper, {the} method and solution used to evaluate and compare the consumption power of the application layer protocols is the method of simulating and modeling the aforementioned protocols in a virtual platform and using special tools that allow the creation and design of the network and provides the intended environment. For this purpose, the Contiki operating system is used, which is specially designed for IoT applications.
In order to simulate and implement the main steps of the research, a simulator that is available in the Contiki operating system was used. By means of this simulator, which is called COOJA, it is possible to form the desired network by using the implementations made in the Contiki operating system and run protocols on virtual platforms inside the simulator. To evaluate the power consumption, the powertrace tool is used. Contiki is an operating system for networked and memory-constrained systems with a focus on wireless and low-power devices in the Internet of Things. Existing uses of the Contiki operating system include street lighting systems, sound management and monitoring for smart cities, temperature and alarms. The Contiki operating system is equipped with a sensor simulator called Cooja, which can be used to simulate the nodes in Contiki. Contiki provides optional multi-threading of proprietary preprocessors. It has also provided the possibility of inter-process communication by sending messages between events.

COOJA simulator supports many application platforms. These platforms are ready-to-use in this simulator, and there is no need to install or add a special program or plugin to use them. Among the platforms supported by COOJA are TelosB/SkyMote, Zolertia Z1, Wismote, ESB and MicaZ. In this paper, the Zolertia Z1 platform is used, which can be found in the COOJA simulator named Z1 mote. This module has a low-power MSP430F2617 microcontroller with a frequency of 16 MHz. It also has a RAM of 8 KB and a flash memory of 92 KB. This module has a CC2420 transmitter with a speed of 250 Kbps and a frequency of 2.4 GHz. The sensors in this module include a temperature sensor and a three-axis digital accelerometer. It also has a built-in ceramic antenna. It should be noted that the clock speed in the Z1 module defined in COOJA is 8 MHz, while the actual module has a clock speed of 16 MHz.

To calculate and analyze the power consumption of the protocols, Powertrace is used, which is a program written in C language. By running this program in COOJA, in the output of each node, the power consumption module can be seen in several stages of the modules under the title ALL\_CPU, ALL\_LPM, ALL\_TX and ALL\_RX. ALL\_CPU is actually the sum of the clock ticks in the state when the CPU is active. ALL\_LPM is the set of ticks when the LPM mode is active or in other words the CPU is on low power mode. ALL\_TX and ALL\_RX are respectively the number of clock ticks in the mode of sending and receiving information and packets. These outputs are all accessible by enabling the Powertrace tool that Contiki's operating system allows to use. The consumed power is evaluated based on Equation \ref{Eq1} \cite{ref24}.
\begin{small}
\begin{equation}
Power \ consumption =\frac{Energyest_{value}*Current*Volatge}{RTimer_{second}*Runtime}
\label{Eq1}
\end{equation}
\end{small}
In Eq. \ref{Eq1}, the value of Energyest is the number of clock ticks counted by Powertrace between two time intervals, which are in one of the CPU, LPM, TX or RX modes. In fact, these values are the main values that have the extension ALL in the main code of the Powertrace program. After that, we reach the amount of electric current, which in this paper, since we have used Z1 as a node in the simulation, its electric current can be extracted from the datasheet of Z1 module.
It is important to note that the electric current of the modules used in each of its working modes is different and should not be considered the same. According to its manual, the voltage of Z1 is equal to 3V and {RTimer} is equal to 32768. The execution time is also the time interval between two samplings in the simulation, which in this paper, the time interval is 10 seconds.

Contiki operating system is used to receive data collected through sensors, using sensor DDS software, MQTT protocol and MQTT-SN protocol. The Energyest module is also used in the Contiki operating system to calculate and evaluate the power consumption of devices that are limited in terms of hardware resources.
The evaluation of power consumption can also be done using the macros given in the following Equations:
\begin{small}
\begin{equation}
Power_{CPU} =\frac{Energyest_{CPU}*Current*Volatge}{RTimer_{second}*Runtime}
\label{Eq2}
\end{equation}
\end{small}
\begin{small}
\begin{equation}
Power_{LPM} =\frac{Energyest_{LPM}*Current*Volatge}{RTimer_{second}*Runtime}
\label{Eq3}
\end{equation}
\end{small}
\begin{small}
\begin{equation}
Power_{Transmitter} =\frac{Energyest_{Transmitter}*Current*Volatge}{RTimer_{second}*Runtime}
\label{Eq4}
\end{equation}
\end{small}
\begin{small}
\begin{equation}
Power_{Reciever} =\frac{Energyest_{Reciever}*Current*Volatge}{RTimer_{second}*Runtime}
\label{Eq5}
\end{equation}
\end{small}

In Equations \ref{Eq2}, \ref{Eq3}, \ref{Eq4}, and \ref{Eq5} as already stated, Energyest is actually the number of clock ticks that are in one of the states of CPU, LPM, Transmitter and Receiver. In other words, these values are the same values that the Powertrace program prints every few seconds (in this case, every 10 seconds) on the output of the simulator when the simulation starts. The voltage and current values can also be found in the datasheet of the used device.
In this paper, the Z1 node is used, so you can find the voltage and current in its catalog. The voltage value to be used in the 3V and the current value are also considered according to the datasheet of the Z1 module in each case, which is largely close to the average values used. The RTIMER value is a fixed value equal to 32768 and the execution time, which can be called Interval Time, is set to 10 seconds in this paper. This whole interval or Interval Time in this test is equal to 100 seconds.
The point that should be taken into account in the meantime is that the above values and equations are used to obtain the power consumption in each of the situations, and to reach the final value for the calculation of the power consumption, all the obtained values must be added together as shown in Equation \ref{Eq6}:
\begin{small}
\begin{equation}
\begin{split}
Power_{Total} & =Power_{CPU}+Power_{LPM}\\
 &+Power_{Tranasmitter}+Power_{Reciever}
\end {split}
\label{Eq6}
\end{equation}
\end{small}
To obtain the power of the battery used for the devices, it is enough to use Equation \ref{Eq7}.
\begin{small}
\begin{equation}
Power_{Battry} =Current*Voltage
\label{Eq7}
\end{equation}
\end{small}

According to the Equation \ref{Eq7}, the battery power used for devices is calculated to be 7.5W.

\section{Evaluation Results}
\label{results}

\subsection{Simulation parameters}
The simulation parameters are presented in Table \ref{tab1}. 

\begin{table}
\caption{Simulation parameters.}
\begin{tabular}{|c|c|}
\hline 
Parameters &  Values\\ 
\hline 
Operating system & Instant Contiki 3.0 \\ 
\hline 
Simulator & COOJA \\ 
\hline 
Application layer protocols & MQTT, MQTT-SN, CoAP, HTTP \\ 
\hline 
Node type & Zolertia Z1 mote \\ \hline
IP protocol version & IPv6\\ \hline
Radio type &Unit Disk Graph Medium (UDGM) \\ \hline
Number of clients &1 \\  \hline
Simulation time &100 sec \\ \hline

\hline 
\end{tabular} 
\label{tab1}
\end{table}

\subsection{MQTT evaluation}

In this simulation, the broker used is Really Small Message Broker (RSMB), which, as its name suggests, is a light-weight message broker with very low overhead that can send messages from/to small devices such as sensors and arms on a network with limited bandwidth and processing capabilities, etc. This broker works like traditional MQTT brokers such as Mosquitto, but the main difference is that for RSMB, a bridge needs to be created to establish a connection with the broker. Table \ref{tab2} shows the measured power consumption for the MQTT protocol, calculated in milliwatt hours (mWh).
\begin{table}
\caption{MQTT power consumption. }
\resizebox{0.5\textwidth}{!}{ 
\begin{tabular}{|c|c|c|c|c|c|}
\hline 
Time &CPU & LPM & TX & RX & Total \\ 
\hline 
10 & 0.288253784 & 0.000294056 & 0.269165039 & 0.404983521 & 0.9626964 \\ 
\hline 
20 & 0.259918213 & 0.000294674 & 0.650024414 & 0.4497789 & 1.360016201 \\ 
\hline 
30 & 0.261383057 & 0.00029464 & 0.389099121 & 0.449780273 & 1.100557091 \\ 
\hline 
40 & 0.126159668 & 0.00029735 & 0.190429688 & 0.449890137 & 0.766776843 \\ 
\hline 
50 & 0.125518799 & 0.000297361 & 0.190429688 & 0.44989151 & 0.766137357 \\ 
\hline 
60 &0.107803345 & 0.00029314 & 0.190429688 & 0.44989151 & 0.748417682 \\ 
\hline 
70 & 0.133346558 & 0.000297203 & 0.264587402 & 0.4497789 & 0.848010063 \\ 
\hline 
80 & 0.123275757 &0.000297404 & 0.194091797 & 0.449890137 & 0.767555095 \\ 
\hline 
90 & 0.125930786 & 0.000297353 & 0.57220459 & 0.449890137 & 1.148322866 \\ 
\hline 
100& 0.131561279 & 0.000297241 & 0.839538574 & 0.449785767 & 1.421182861 \\ 
\hline 
Average & 0.168315125 &0.000296042 & 0.375 & 0.445356079 &0.988967246 \\ 
\hline 
\end{tabular} }
\label{tab2}
\end{table}

In Table \ref{tab2}, the power consumption in different modes such as CPU, LPM, Transmit and Receive is calculated separately, and finally, by summing the power consumption in each mode, the values of the Total column are obtained. In the last row of the table, the average power in each mode is calculated separately and finally the total average.
In the CPU mode in the MQTT protocol, initially, more energy is always spent communicating with the client. In this case, the highest amount of consumption is related to the time of 10 seconds, which is equal to 0.288253574 mW, and the lowest consumption is related to the time of 60 seconds, which is 0.107803345 mW. The average consumption in this case is equal to 0.168315125 mW.
  In low consumption mode or LPM, the lowest energy consumption is recorded in 60 seconds {with 0.00029314 mW, the highest consumption is recorded in 80 seconds with 0.000297404 mW,} and the average of this mode is 0.000296042 mW.
In TX mode, the highest consumption corresponds to the time of 100 seconds with a value of 0.839538574 mW and the lowest consumption corresponds to the times of 40, 50 and 60 seconds with values equal to 0.190429688 mW. The average consumption in this case is equal to 0.375 mW. In the column related to RX, which on average causes the highest power consumption in the MQTT protocol, the lowest consumption in 10 seconds is 0.404983521 mW and the highest consumption is in 50 and 60 seconds with a value of 0.44989151 mW, which is the average power consumption in this case is equal to 0.445356079 mW.

\subsection{MQTT-SN evaluation}
In the MQTT-SN protocol simulation, the RSMB broker is used as before and the broker\_mqtts broker is configured as a part of the RSMB.
 Table \ref{tab3} shows the measured power consumption for the MQTT protocol, calculated in milliwatt hours (mWh).
\begin{table}
\caption{ MQTT-SN power consumption.}
\resizebox{0.5\textwidth}{!}{ 
\begin{tabular}{|c|c|c|c|c|c|}
\hline 
Time &CPU & LPM & TX & RX & Total \\ 
\hline 
10 & 0.30368042 & 0.000293752 & 0.205993652 & 0.40488739 & 0.914855214 \\ 
\hline 
20 & 0.375915527 & 0.000292354 & 0.14465332 & 0.44901947 & 0.969880672 \\ 
\hline 
30 & 0.322311401 & 0.00029343 & 0.14465332 & 0.449408112 & 0.916666263 \\ 
\hline 
40 & 0.150558472 & 0.00029687 & 0.071411133 & 0.449710236 & 0.67197671 \\ 
\hline 
50 & 0.150604248 & 0.000296869 & 0.070495605 & 0.449710236 & 0.671106958 \\ 
\hline 
60 &0.278137207 & 0.000294316 & 0.070495605 & 0.449714355 & 0.798641484 \\ 
\hline 
70 & 0.278137207 & 0.000296708 & 0.14465332 & 0.449598999 & 0.752981034 \\ 
\hline 
80 &0.197433472 &0.000295935 & 0.070495605 & 0.449704742 & 0.717929755 \\ 
\hline 
90 & 0.251358032 & 0.000294856 & 0.070495605 & 0.4491362 & 0.771284693 \\ 
\hline 
100& 0.235336304 & 0.00029517 & 0.140991211 & 0.448726959 & 0.825349644 \\ 
\hline 
Average & 0.242376709 &0.000295026 & 0.113433838 & 0.44496167 &0.801067243 \\ 
\hline 
\end{tabular} }
\label{tab3}
\end{table}
It should be noted that like the MQTT protocol, at the beginning of communication, more power is consumed than at other times, the average consumption in CPU mode is equal to 0.242376709 mW, which is 0.08 mW more than the average consumption in CPU mode of the MQTT protocol. In LPM mode, it had the highest consumption in 50 seconds with 0.000296869 mW and the lowest consumption in 20 seconds with 0.000292354 mW. The average power consumption in this case is equal to 0.000295026 mW. In TX mode, the highest consumption is related to the time of 10 seconds with the amount of consumption of 0.205993652 mW and the lowest consumption is related to the times of 50, 60, 80 and 90 seconds, in which the amount of power consumption is equal to 0.070495605 mW. The average consumption in this case is 0.113433838 mW. In RX mode, the highest consumption was in 60 seconds with a value of 0.449714355 mW, and the lowest consumption was in 10 seconds with a value of 0.40488739 mW. The average power consumption in TX mode was equal to 0.44496167 mW, which is the highest power consumption in this mode. On average, this protocol consumed a total of 0.801067243 mW. So far, a significant reduction in power consumption can be seen in the MQTT-SN protocol compared to MQTT. Power reduction in TX mode is the main factor in low power consumption of MQTT-SN.

\subsection{COAP evaluation}
The general process of implementing this protocol in this project is completely similar to MQTT and MQTT-SN, with the difference that there is no need to build a bridge and RSMB broker, {it is enough to add the codes related to powertrace to the simulation executable file and make the necessary changes to the main code of the program.}
Table \ref{tab4} shows the measured power consumption for the COAP protocol, calculated in milliwatt hours (mWh).

\begin{table}
\caption{ COAP power consumption.}
\resizebox{0.5\textwidth}{!}{ 
\begin{tabular}{|c|c|c|c|c|c|}
\hline 
Time &CPU & LPM & TX & RX & Total \\ 
\hline 
10 & 0.2416752 & 0.00029156 & 0.06914837 & 0.41543256 & 0.72654769 \\ 
\hline 
20 & 0.25167342 & 0.00029532 & 0.08924518 & 0.42516849 & 0.425168490.76638241 \\ 
\hline 
30 & 0.32154684 & 0.00029354 & 0.14189535 & 0.53545168 & 0.99918741 \\ 
\hline 
40 & 0.27301025 & 0.00029446 & 0.08513495 & 0.46158435 & 0.82002401 \\ 
\hline 
50 &0.34156846 & 0.00029021 & 0.35324895 & 0.55012351 & 1.24523113 \\ 
\hline 
60 &0.22134692 & 0.00029254 & 0.07125146 & 0.45156578 & 0.84365581 \\ 
\hline 
70 & 0.21034297 & 0.00029394 &0.08015423 & 0.46157525 & 0.75236639 \\ 
\hline 
80 &0.20026484 &0.00029154 & 0.09215432 & 0.45154862 & 0.74425932 \\ 
\hline 
90 & 0.41354568 & 0.00029121 & 0.31514925 & 0.52481235 & 1.52528289 \\ 
\hline 
100& 0.28689859 & 0.00029511 & 0.099981 & 0.48156584 & 0.86874054 \\ 
\hline 
Average & 0.276187317 &0.000292943 & 0.139736306 & 0.475882843 &0.92916776 \\ 
\hline 
\end{tabular} }
\label{tab4}
\end{table}

In this table, in the first column, which is related to the power consumption in the processing mode, the highest power is seen in 90 seconds with a value of 0.41354568 mW, and the lowest consumption is related to the time of 80 seconds with a power consumption of 0.20026484 mW. The average power consumption is equal to 0.276187317 mW. In total, the highest consumption in 90 seconds with a value of 1.52528289 mW and the lowest total consumption in 10 seconds with a value of 0.72654769 mW and the average total power consumption for the CoAP protocol is equal to 0.92916776 mW, which is slightly less than the MQTT protocol and a bit more than the MQTT-SN protocol.
\subsection{HTTP evaluation}
Like CoAP, the HTTP protocol does not need to run and launch the RSMB broker like the other two protocols, namely MQTT and MQTT-SN, and it can be run by a client after creating a simulation space for HTTP, and the simulation results can be extracted by powertrace.
Table \ref{tab5} shows the measured power consumption for the {HTTP} protocol, calculated in milliwatt hours (mWh).

\begin{table}
\caption{ HTTP power consumption.}
\resizebox{0.5\textwidth}{!}{ 
\begin{tabular}{|c|c|c|c|c|c|}
\hline 
Time &CPU & LPM & TX & RX & Total \\ 
\hline 
10 & 0.253967285 & 0.000289693 & 0.013549805 &0.352478027 & 0.62028481 \\ 
\hline 
20 & 0.168983459 & 0.00029313 & 0 & 0.741485596 &0.910762185 \\ 
\hline 
30 & 0.157424927 & 0.000293596 & 0.289764404 & 0.707839966 &1.155322893 \\ 
\hline 
40 & 0.060997009 & 0.000297451 & 0.532653809 & 0.658172607 & 1.252120876 \\ 
\hline 
50 &0.050262451 & 0.000297881 & 0.282348633 & 0.615234375 & 0.94814334 \\ 
\hline 
60 &0.156761169 &0.000293627 & 0 & 0.669708252 & 0.826763048 \\ 
\hline 
70 & 0.547622681 &0.000277944&1.076751709 & 1.370819092 & 2.995471425 \\ 
\hline 
80 &0.057861328 &0.000297563 & 1.161804199 & 0.615234375 &1.835197465 \\ 
\hline 
90 & 0.052940369 & 0.000297766 & 0.629425049 & 0.615234375 & 1.297897559 \\ 
\hline 
100& 0.055755615 &0.000297653 & 1.247314453 & 0.691177368 & 1.994545089 \\ 
\hline 
Average & 0.156257629 &0.00029363 &0.523361206 &0.703738403 &1.383650869 \\ 
\hline 
\end{tabular} }
\label{tab5}
\end{table}
In this table, which has many differences in terms of values from the previous tables, in processing mode, in the worst case, the power consumption is 0.547622681 mW in 70 seconds, and in 50 seconds, we have the lowest power with a value of 0.050262451 mW. In this mode, the average power consumption is equal to 0.156257629 mW. In LPM mode, it is not much different from other protocols. The highest power consumption in 50 seconds is equal to 0.000297881 mW and the lowest consumption in 20 seconds is 0.00029313 mW. On average, the power consumption of LPM mode is equal to 0.00029363 mW. In TX mode, HTTP protocol consumes a lot of power. In the highest mode in 100 seconds, the power consumption is equal to 1.247314453 mW, and in the lowest mode, it did not occur in the 20th and 60th transmission times, so the power consumption is 0 mW, and on average, in TX mode, the power consumption is calculated as 0.523341206 mW.
In total, the HTTP protocol has an average power consumption of 1.383650869 mW, which is more than all other protocols. This shows that the HTTP protocol is not suitable for use in IoT, where most of the devices used have limitations and require the use of low-power generators, which causes the battery life of the devices to be lower than usual. It also reduces the lifespan of devices and sensors.

\subsection{Analyzing the results and comparing the performance of the protocols }

\begin{figure}[htbp]
\centerline{\includegraphics[scale=0.5]{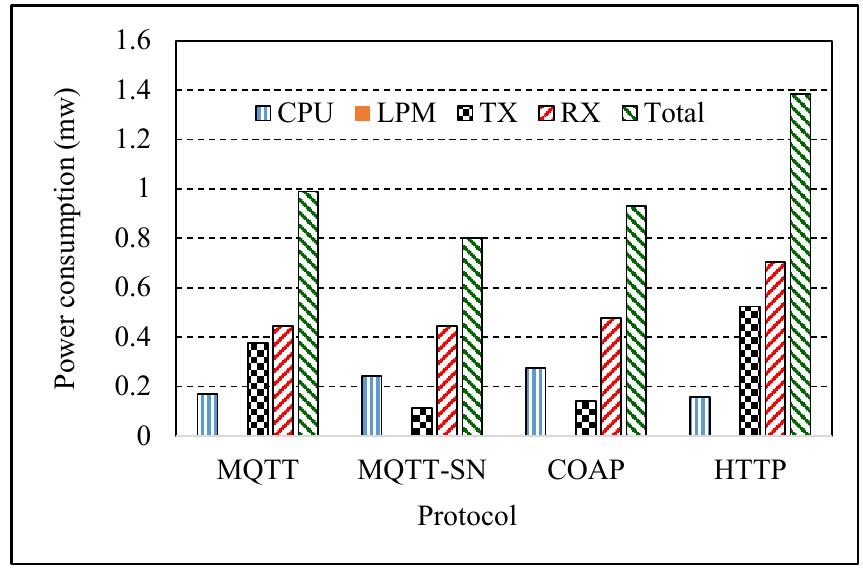}}
\caption{Power evaluation of protocols.}
\vspace{-0.25cm}
\label{fig2}
\end{figure}

After extracting the required results from the simulator and observing the tables related to the consumption of each protocol, as well as observing the average consumption of each of them in different modes and in the total mode, it can be generally concluded that the HTTP protocol is the most consuming and the heaviest protocol. MQTT-SN is the lightest protocol among the simulated ones.
Figure \ref{fig2} depicts the average power consumption of different protocols.

It can be seen from the graph that the power consumed in LPM mode does not contribute much to the overall power consumption of the protocols and it is always a small amount which can even be ignored in this case and by testing with a client.
But on the other hand, the average power consumption in RX mode in all protocols is smooth and significant, which has the highest consumption in MQTT, MQTT-SN and CoAP protocols. Of course, the same thing applies to the HTTP protocol, but the problem here is that in this protocol, the average power consumption in TX mode is also very high, but at the same time, it has a lower consumption in processing mode than other protocols. According to the graph, MQTT-SN protocol consumes less power than other protocols, even CoAP.
In the comparison of two protocols, CoAP and MQTT, it can be seen that they consume somewhat the same amount in RX mode, but in CPU mode, CoAP protocol has a higher consumption, while in TX, MQTT protocol consumes much more, so that in the end, the total consumption power of MQTT protocol is slightly higher than CoAP.

Protocols have different consumption power in different modes. For example, the MQTT-SN protocol consumes about 30\% more power than MQTT in CPU mode, but due to the more optimal MQTT-SN protocol and 70\% reduction in power consumption in TX mode, in general and in total, consumption is 19\% compared to The MQTT protocol has been deprecated.

Comparing CoAP and MQTT, it can be seen that the consumption of CoAP protocol is about 40\% more in CPU mode. In RX mode, CoAP has a higher consumption of about 7\%, but the main difference in power consumption between these two protocols is evident in TX mode. In this mode, the CoAP protocol is about 67\% less consuming than MQTT, and in the total consumption of the two protocols, CoAP is 6\% more optimal than MQTT protocol in terms of power.

\section{Conclusion}
\label{conclusion}
In this paper, based on the obtained results, the optimality of the discussed protocols in use for communication in the field of Internet of Things was analyzed and investigated.
In general and all modes, the MQTT-SN protocol is the least consumed protocol among them. Then the CoAP protocol and then the MQTT protocol, which has a slightly higher power consumption than CoAP. Finally, the HTTP protocol is placed, which consumes very high power in the same time compared to other protocols, so that the wide use of this protocol in communication is not recommended at all in the field of Internet of Things. Therefore, using these results, it is possible to use protocols and connections that have the lowest power consumption and the highest efficiency and speed to improve the performance of the devices, increase battery life and reduce depreciation and increase the life of the devices used.

\bibliographystyle{IEEEtran}
\bibliography{ref}

\end{document}